\documentclass[12pt, fleqn]{article}
\usepackage[cp1251]{inputenc}
\usepackage{latexsym,amsfonts,amssymb}
\usepackage{graphicx}

\usepackage{amsbsy}
\usepackage{amsmath}
\usepackage{epsf}

\newtheorem{theo}{Theorem}
\newcommand{\bt}{\begin{theo}}
\newcommand{\et}{\end{theo}}
\newcommand{\bd}{\begin{displaymath}}
\newcommand{\ed}{\end{displaymath}}

\newcommand{\be} {\begin{equation}}
\newcommand{\ee} {\end{equation}}
\newcommand{\ba} {\begin{array}}
\newcommand{\ea} {\end{array}}

\newcommand{\p} {\partial}

\begin{document}
 \begin{center}
 {\Large \bf Reaction-diffusion systems with constant diffusivities: conditional symmetries and \\ form-preserving transformations}\\
 \medskip

{\bf Roman Cherniha$^{\dag,\ddag}$}
 {\bf and  Vasyl' Davydovych$^\dag$}
 \\
{\it  $^\dag$~Institute of Mathematics, Ukrainian National Academy
of Sciences,\\
 3 Tereshchenkivs'ka Street, Kyiv 01601, Ukraine\\
  $^\ddag$~Department  of  Mathematics,
     National University
     `Kyiv-Mohyla Academy', \\ 2 Skovoroda Street,
     Kyiv  04070 ,  Ukraine
 }\\
 \medskip
 E-mail: cherniha@imath.kiev.ua and davydovych@imath.kiev.ua
\end{center}

\begin{abstract}

 $Q$-conditional symmetries (nonclassical symmetries)
 for a general  class of two-component reaction-diffusion   systems  with constant diffusivities  are studied.
 Using the  recently  introduced  notion of  $Q$-conditional symmetries
 of the first type (R. Cherniha  J. Phys. A: Math.
Theor., 2010. vol.~43., 405207), an  exhaustive  list of
reaction-diffusion  systems  admitting  such symmetry is derived.
 The form-preserving transformations for  this class of
 systems  are constructed and it is shown that  this  list  contains only  non-equivalent  systems.
  The obtained symmetries permit to  reduce the reaction-diffusion   systems under study to two-dimensional   systems of ordinary differential equations   and  to find exact solutions.
 As a non-trivial example,  multiparameter  families  of exact solutions are explicitly  constructed for two  nonlinear   reaction-diffusion  systems.
 A possible   interpretation  to  a  biologically  motivated model
   is presented.

\end{abstract}

\section{\bf Introduction}
 The  paper is devoted  to the  investigation of
 the two-component reaction-diffusion (RD) systems of the form
 \begin{equation}\label{1}\begin{array}{l}
u_t=d_1u_{xx}+F(u,v), \\
v_t=d_2v_{xx}+G(u,v).\end{array}\end{equation} where
  $u=u(t,x)$ and $v=v(t,x)$ are two  unknown functions representing   the densities
  of populations (cells), the concentrations of chemicals, the pressures in  thin
films, etc. $F$ and $G$ are the  given smooth functions describing
interaction between them and environment,
 $d_1$ and $d_2$ are diffusivities assumed to be positive constants.  The subscripts $t$ and $x$ denote
differentiation with respect to these variables. The class of  RD
systems  (\ref{1}) generalizes many well-known nonlinear
second-order models and is used to describe various processes in
physics, biology, chemistry  and ecology (see, e.g., the well-known
books
 \cite{ames, mur2, mur2003, aris, okubo}).

Nevertheless the search  for  Lie symmetries of  the class of RD
systems  (\ref{1}) was initiated about 30 years ago
\cite{zulehner-ames}, this problem was  completely solved only
during the last decade   because of its complexity.
 Now
one can claim that all possible  Lie symmetries of  (\ref{1})
 were  completely described
in \cite{ch-king,ch-king2, niki-05}.

The time is therefore ripe for a complete description
 of non-Lie symmetries for  the class of the RD systems  (\ref{1}).
However,  it  seems to be extremely difficult task because, firstly,
 several definitions of non-Lie symmetries have been introduced
 (nonclassical symmetry \cite{ ames,bl-c}, conditional symmetry \cite{Fush93, ch-he-2004},
  generalized conditional symmetry
\cite{Foka94,liu-fokas} etc.), secondly, the complete description
 of non-Lie symmetries  needs to solve the corresponding
 system of determining equations, which is {\it non-linear}
 and can  fully be  solved only in exceptional cases.

 Hereafter we use the most common  notion  among  non-Lie symmetries, non-classical symmetry, which we continuously call the $Q$-conditional symmetry
following the well-known book \cite{Fush93} and our previous papers
\cite{ch-dav-2011,ch-dav-2012}.
 It is  well-known that  the notion of $Q$-conditional symmetry plays
an important role in investigation of the nonlinear evolution
equations
because,  having such symmetries in the explicit form, one may
construct new exact solutions, which are not  obtainable  by the
classical Lie machinery. However, for a complete description
 of  such  symmetries, one   needs to solve the corresponding
non-linear  system of determining equations that usually
 is very difficult task.
Thus,
  to solve the $Q$-conditional symmetry  classification problem
 for  the class of RD systems  (\ref{1}), one should look for  new constructive approaches helping to  solve  the relevant  nonlinear system of determining equations.
 A possible approach was recently proposed in   \cite{ch-2010}  and   is used  in this paper.

It can be noted  that the diffusion coefficient $d_1$ in  system
(\ref{1}) can be omitted without losing of generality because the
simple substitution \begin{equation}\nonumber t\rightarrow t/d_1,  F
\to -d_1C^1, G \to -d_2C^2\end{equation}
 reduces the system to the form
  \begin{equation}\label{3}\begin{array}{l}
u_{xx}=u_t+C^1(u,v),
\\v_{xx}=dv_t+C^2(u,v),\end{array}\end{equation}  where $d=\frac{d_1}{d_2}.$  Thus, we consider system  (\ref{3}) in what follows.

The paper is organized as follows.
  In  section 2,   two different  definitions of   $Q$-conditional
   invariance for the class of RD systems   (\ref{3}) are presented and  the system of   determining equations is derived.
  The  theorem  giving the complete description of $Q$-conditional symmetries of the first  type   is proved.
In section 3, the form-preserving transformations for  the class of
RD systems  (\ref{3}) are constructed  and applied to the RD systems
derived in section 2.
  In section 4,  the $Q$-conditional  symmetry   obtained for
   reducing of   the RD systems to the ODE systems   are applied.
   Examples  of finding
  exact solutions are presented together with a possible  interpretation for population dynamics.
   Finally, we  summarize and discuss
   the results obtained   in
the last section.
\section{\bf Conditional symmetries of the RD systems}
Here   we use the definition of $Q$-conditional symmetry of the
first type for  the RD systems (see \cite{ch-2010} for details). It
is well-known that to find Lie invariance  operators, one needs to
consider   system (\ref{3}) as the manifold
${\cal{M}}=\{S_1=0,\,S_2=0 \}$  where \begin{equation}\nonumber
\begin{array}{l}
 S_1 \equiv \ u_{xx}-u_t-C^1(u,v),\\
S_2 \equiv \ v_{xx}-dv_t-C^2(u,v), \end{array}\end{equation} in the
prolonged space of the  variables: $t, x, u, v, u_t, v_t$,$ u_{x},
v_{x}, u_{xx}, v_{xx}, u_{xt}, v_{xt}, u_{tt}, v_{tt}.$ According to
the definition, system (\ref{3}) is invariant under the
transformations generated by the infinitesimal operator
\begin{equation}\label{2-2} Q = \xi^0 (t, x, u, v)\partial_{t} + \xi^1 (t, x, u,
v)\partial_{x} +
 \eta^1(t, x, u, v)\partial_{u}+\eta^2(t, x, u, v)\partial_{v},  \end{equation}
if the following invariance conditions are satisfied:
\begin{equation}\nonumber \begin{array}{l}
\mbox{\raisebox{-1.6ex}{$\stackrel{\displaystyle Q}{\scriptstyle
2}$}}\, S_1
 \Big\vert_{\cal{M}}=0, \\[0.3cm]
\mbox{\raisebox{-1.6ex}{$\stackrel{\displaystyle Q}{\scriptstyle
2}$}}\, S_2 \Big\vert_{\cal{M}}=0. \end{array} \end{equation}
The operator $ \mbox{\raisebox{-1.6ex}{$\stackrel{\displaystyle  
Q}{\scriptstyle 2}$}} $  
is the second  
 prolongation of the operator $Q$, i.e.  
\begin{equation}\nonumber  
\mbox{\raisebox{-1.6ex}{$\stackrel{\displaystyle  
Q}{\scriptstyle 2}$}}  
 = Q + \rho_t^1\frac{\partial}{\partial u_{t}}+
 \rho_t^2\frac{\partial}{\partial v_{t}}+  
\rho^1_x\frac{\partial}{\partial u_{x}}+ \rho^2_x\frac{\partial}{\partial v_{x}}  
+\sigma_{xx}^1\frac{\partial}{\partial u_{xx}}  
+\sigma_{xx}^2\frac{\partial}{\partial  v_{xx}},  
\end{equation}  
where the coefficients $\rho$ and $\sigma$ with relevant subscripts  
are expressed  via the functions $\xi^0, \xi^1, \eta^1$ and $\eta^2$
by well-known formulae (see, e.g., \cite{Fush93, olv, b-k}).

Hereafter   the listed above differential operators act on functions
and  differential expressions in a natural way, particularly  $\
Q\,(u) =\xi^0u_t+\xi^1u_x-\eta^1 $ and   $ Q\,(v) =
\xi^0v_t+\xi^1v_x-\eta^2$.

\medskip

\noindent \textbf{Definition 1.}\cite{ch-2010}  Operator (\ref{2-2})
is called the $Q$-conditional symmetry of the first type for the RD
system (\ref{3}) if  the following invariance conditions are
satisfied:
\begin{equation}\nonumber \begin{array}{l}
\mbox{\raisebox{-1.6ex}{$\stackrel{\displaystyle Q}{\scriptstyle
2}$}}\, S_1
\Big\vert_{{\cal{M}}_1}=0, \\[0.3cm]
\mbox{\raisebox{-1.6ex}{$\stackrel{\displaystyle Q}{\scriptstyle
2}$}} \,S_2 \Big\vert_{{\cal{M}}_1}=0, \end{array} \end{equation}
where the manifold ${\cal{M}}_1$ is either $\{S_1=0, \,S_2=0,\,
Q\,(u)=0 \}$ or $\{S_1=0,\,S_2=0, \,Q\,(v)=0 \}$.

\medskip

\noindent \textbf{Definition 2.} Operator (\ref{2-2}) is called the
$Q$-conditional symmetry of the second  type, i.e., the standard
non-classical symmetry  for the RD system (\ref{3}) if  the
following invariance conditions are satisfied:
\begin{equation}\nonumber \begin{array}{l}
\mbox{\raisebox{-1.6ex}{$\stackrel{\displaystyle Q}{\scriptstyle
2}$}}\, S_1
 \Big\vert_{{\cal{M}}_2}=0, \\[0.3cm]
\mbox{\raisebox{-1.6ex}{$\stackrel{\displaystyle Q}{\scriptstyle
2}$}}\, S_2
 \Big\vert_{{\cal{M}}_2}=0, \end{array} \end{equation} where the
manifold ${\cal{M}}_2=\{S_1=0,\, S_2=0,\,
Q\,(u)=0, \, Q\,(v)=0 \}$.

\medskip

\noindent \textbf{Remark 1.} It is easily seen that ${\cal{M}}_2
\subset {\cal{M}}_1 \subset {\cal{M}}$, hence, each Lie symmetry is
automatically the  $Q$-conditional symmetry of the first and second
type, while each $Q$-conditional symmetry of the first type is one
of the second type (non-classical symmetry).

\medskip

\noindent \textbf{Remark 2.} To the  best of our  knowledge, there
are not many paper devoted to search of $Q$-conditional symmetries
for {\it the systems of PDEs} \cite{ch-pli-08,  barannyk2002,
 murata-06, arrigo2010,ch-se-03}.
 One may easily check  that Definition 2 was only used in all these  papers.

\medskip

\noindent \textbf{Statement.}  Let us assume that
\begin{equation}\label{3-ad}
X=(h_1(t,x)v+h_0(t,x))\partial_v \end{equation} (hereafter
$h_1(t,x)$ and $h_0(t,x)$ are the given functions)  is the Lie
symmetry operator of the RD system (\ref{3}) while    $Q_1$ is the
known  $Q$-conditional symmetry of the first type, which was found
using  the manifold ${\cal{M}}_1=\{S_1=0, S_2=0, Q\,(u)=0  \}$.
 Then any linear combination
 $C_1Q_1+C_2X$  ($C_1$ and $C_2\not=0$ are arbitrary constants) produces  new  $Q$-conditional symmetry of the first type.

\medskip

Application of   definition 2  for finding Q-conditional symmetry
(non-classical symmetry) operators of  the RD system (\ref{3}) leads
to a complicated system   of determining equations ( DEs) (see
system (19) in \cite{ch-2010}), which seems to be extremely
difficult for exact solving.

It turns out that application of  definition 1 leads to essentially
simpler system of DEs, which can be fully integrated. Here we
present the result under the restrictions  $\xi^0\neq0$ and $d\neq1$
(the cases $\xi^0=0$ and $d=1$ must be
 investigated separately). Thus,  the  system of DEs corresponding to the manifold ${\cal{M}}_1=\{S_1=0,\, S_2=0,\, Q\,(u)=0
 \}$ takes the form
\begin{eqnarray}
 &&
 \xi^0_{x}=\xi^0_{u}=\xi^0_{v} =\xi^1_{u}=\xi^1_{v}=0,\label{4}
 \\ &&
 \eta^1_{v}=\eta^1_{uu}=\eta^2_{uu}=
\eta^2_{vv}=\eta^2_{uv}=0,\label{5}\\ && \label{6}
2\xi^0\eta^2_{xu}+(d-1)\xi^1\eta^2_u=0,
\\ &&
\label{7}  2\eta^1_{xu}+\xi^1_{t}=0,\\ && \label{9}
2\eta^2_{xv}+d\xi^1_t=0,\\ && \label{8} 2\xi^1_x-\xi^0_t =0,\\ &&
\label{10} \eta^1C^1_u+\eta^2C^1_v+(2\xi^1_x-
\eta^1_u)C^1=\eta^1_{xx}-\eta^1_t,\\ && \label{11}
\eta^1C^2_u+\eta^2C^2_v+ (2\xi^1_x-\eta^2_v)C^2=\eta^2_uC^1+ (1-
d)\frac{\eta^1}{\xi^0}\eta^2_u+\eta^2_{xx}-d\eta^2_t.
\end{eqnarray}

Note that  there is no any need to solve  the similar system  of DEs
corresponding to the manifold ${\cal{M}}^*_1=\{S_1=0, S_2=0,
Q\,(v)=0
 \}$ because the discrete transformations
 $u\rightarrow v, \, v\rightarrow
 u$  transform each symmetry found using  ${\cal{M}}_1$ to one   corresponding to the manifold  ${\cal{M}}^*_1$.

 It should be also noted that we find purely conditional symmetry operators, i.e., exclude all
  such operators, which are equivalent to Lie symmetry operators described in  \cite{ch-king, ch-king2}.
Having  this aim,  we use the system DEs for search Lie symmetry
operators (see \cite{ch-dav-2012} for details):
\begin{eqnarray}
 &&
\label{112}
 \xi^0_{x}=\xi^0_{u}=\xi^0_{v} =\xi^1_{u}=\xi^1_{v}=0,
\\&&
\label{113} \eta^1_{v}=\eta^2_{u}=\eta^1_{uu}=
\eta^2_{vv}=0,\\&&\label{114} 2\xi^1_x-\xi^0_t=0,\\&&\label{116}
2\eta^1_{xu}+\xi^1_{t}=0,\\&&\label{117}
2\eta^2_{xv}+d\xi^1_t=0,\\&&\label{118}
\eta^1C^1_u+\eta^2C^1_v+(2\xi^1_x-\eta^1_u)C^1=\eta^1_{xx}-
\eta^1_t,\\&&\label{119} \eta^1C^2_u+\eta^2C^2_v+
(2\xi^1_x-\eta^2_v)C^2=\eta^2_{xx}-d\eta^2_t.
\end{eqnarray}
Comparing  DEs (\ref{4})--(\ref{11}) with (\ref{112})--(\ref{119})
 one concludes that
  $\eta^2_{u}\neq0$ is the necessary and sufficient condition, which guarantees this property.

 Now we need to solve the nonlinear system (\ref{4})--(\ref{11}).
Obviously equations  (\ref{4}) and   (\ref{5}) can be easily
 integrated:\\
 \begin{equation}\label{2-6}\begin{array}{l}  \xi^0=\xi^0(t), \ \xi^1=\xi^1(t,x), \\ \eta^1=r^1(t,x)u+p^1(t,x), \
\eta^2=q(t,x)u+r^2(t,x)v+p^2(t,x), \end{array}\end{equation} where
$\xi^0(t), \, \xi^1(t,x), \, q(t,x), \, r^k(t,x), \, p^k(t,x) $
($k=1, 2$) are
to-be-determined functions. Thus, substituting   (\ref{2-6}) into
(\ref{6})--(\ref{11}), one obtains the nonlinear system of PDEs:
\begin{eqnarray}
 &&
\label{12} \ 2\xi^0q_x+\xi^1(d-1)q=0,\\ && \label{13} \
2r^1_{x}+\xi^1_{t}=0,\\ && \label{14} \ 2r^2_{x}+d\xi^1_t=0,\\
&& \label{15} \ 2\xi^1_x-\xi^0_t =0,\\ && \label{16} \
(r^1u+p^1)C^1_u+(qu+r^2v+p^2)C^1_v+(2\xi^1_x-r^1)C^1=
(r^1_{xx}-r^1_t)u+ p^1_{xx}-p^1_t,\\ && \label{17}
(r^1u+p^1)C^2_u+(qu+r^2v+p^2)C^2_v+
(2\xi^1_x-r^2)C^2=qC^1+\frac{r^1u+p^1}{\xi^0}q(1-d)+\\ &&
(r^2_{xx}-dr^2_t)v+ (q_{xx}-dq_t)u+p^2_{xx}-dp^2_t,\nonumber
\end{eqnarray} to find
the functions
 $\xi^0(t), \, \xi^1(t,x), \, q(t,x)\neq0, \,
r^k(t,x), \, p^k(t,x).$ In other words, all possible $Q$-conditional
symmetries  of the first type are easily constructed provided  the
general solution of system (\ref{12})--(\ref{17}) is known.

\bt The nonlinear RD system (\ref{3}) with \ $d\neq1$ is invariant
under the $Q$-conditional operator of the first type (\ref{2-2}) if
and only if   one  and the corresponding operator have the forms
listed in Table~1. Any other RD system admitting such kind of 
$Q$-conditional operator is reduced to one of those from Table~1 by
the local transformations \begin{equation}\label{32}\begin{array}{l}t\rightarrow C_1t+C_2,\\ x\rightarrow C_3x+C_4, \\
u\rightarrow C_5e^{C_6t}u+C_7t+C_8,\\ v\rightarrow
C_9e^{C_{10}t}v+C_{11}t^2+C_{12}t+C_{13},
\end{array}\end{equation} with correctly-specified constants $C_l,
l=1,\dots,13$ and/or the discrete transformations 
\be\label{d52} u \rightarrow v, \ v
\rightarrow u.\ee Simultaneously the relevant operator is  reduced
by possible adding  a Lie symmetry operator of the form
$(h_1(t,x)v+h_0(t,x))\partial_v$ to those from  Table~1.
 \et
\textbf{ Sketch of  proof.}
 To prove   the theorem  one needs to solve   the nonlinear PDE
system (\ref{12})--(\ref{17}) with restriction  $q(t,x)\neq0$.  We
remind the reader that $C^1$ and  $C^2$ should be  treated as
unknown functions. As
 follows from the preliminary analysis (see  equations (\ref{16}) and  (\ref{17}) involving  the functions $C^1$ and  $C^2$ ), we
should  examine  6 cases:\\
$(1)\ r^1=r^2=p^1=0,$\\
$(2)\ r^1=r^2=0,\ p^1\ne0,$\\
$(3)\ r^1=p^1=0,\ r^2\ne0,$\\
$(4)\ r^1=0,\ p^1\ne0,\ r^2\ne0,$\\
$(5)\ r^2=0,\ r^1\ne0,$ \\
$(6)\ r^1\ne0, \ r^2\ne0.$ \\
Solving  system (\ref{12})--(\ref{17})  in each case one obtains the
list of $Q$-conditional symmetries  of the first type together with
the correctly-specified  functions $C^1$ and  $C^2$. Note that the
symmetry operators have the different structures depending   on the
case.

Let   us consider case {\it (1)}  in details. Equations  (\ref{16})
and  (\ref{17}) take the form
 \begin{equation}\label{27}\begin{array}{l}
(qu+p^2)C^1_v+2\xi^1_xC^1=0,\\
(qu+p^2)C^2_v+2\xi^1_xC^2=qC^1+(q_{xx}-dq_t)u+p^2_{xx}-dp^2_t.\end{array}\end{equation}

Differentiating the first equation of (\ref{27}) with respect to
$x$, one arrives at the equation
 $(q_xu+p^2_x)C^1_v=0,$  which lead to the requirement
$C^1_v=0.$ In fact, if  $q_x\neq0$  then immediately  $C^1_v=0.$ If
$q_x=0$ then equation  (\ref{12})   produces $\xi^1=0$, hence,
$C^1_v=0.$ Thus, the first equation of  system (\ref{27})  takes the
form $\xi^1_xC^1=0$  and two subcases  $\xi^1_x\neq0$ and
$\xi^1_x=0$  should be examined.

The general solution of (\ref{27})  with  $\xi^1_x\neq0$  is
\begin{equation}\label{27**} C^1=0, \
C^2=\exp\big(-\frac{2\xi^1_x}{qu+p^2}v\big)g(u)+\frac{q_{xx}-dq_t}{2\xi^1_x}u+
\frac{p^2_{xx}-dp^2_t}{2\xi^1_x},\end{equation}  where  $g(u)$  is
an arbitrary (at the moment) function. Because  the function $C^2$
doesn't depend on  $t$  and $x$,  equation  (\ref{27**})  with
$g(u)\neq0$   immediately produces the restrictions
$q=\alpha_1\xi^1_x, \, p^2=\alpha_2\xi^1_x,$  where  $\alpha_1$  and
$\alpha_2$  are arbitrary constants.  Differentiating  equation
(\ref{15}) with respect to  $x$,  one obtains $\xi^1_{xx}=0$. So
$q_x\equiv\alpha_1\xi^1_{xx}=0$,  however, this contradicts to the
assumption  $\xi^1_x\neq0$. The remaining possibility  $g(u)=0$
leads to the linear RD system (\ref{3}).

Now we examine the subcase $\xi^1_x=0$, i.e.,
$\xi^1=\lambda_1=const$. The general solution of (\ref{27})    takes
the form
 \begin{equation}\label{27*} C^1=f(u),\ C^2=\frac{qf(u)+(q_{xx}-dq_t)u+p^2_{xx}-dp^2_t}{qu+p^2}v+g(u),\end{equation}
where  $f(u)$ and $ g(u)$  are arbitrary (at the moment) functions.

 If  $f(u)$  is an arbitrary function then  we obtain $p^2=\beta q$ ($\beta =const$ )
Hence $C^2=\frac{f(u)}{u+\beta}v+\alpha v+g(u),$  where
$\alpha=\frac{q_{xx}-dq_t}{q}.$ Having this, we use renaming
$\frac{f(u)}{u+\beta} \to f(u)$  and solve the overdetermined system
  \begin{equation}\nonumber\begin{array}{l} \frac{q_{xx}-dq_t}{q}=\alpha, \\
2q_x+\lambda_1(d-1)q=0. \end{array}\end{equation} Thus, the system
of DEs (\ref{12})--(\ref{17}) is completely solved (under above
listed restrictions !) and we obtain the conditional symmetry
operator
  \begin{equation}\nonumber Q=\partial_t+\lambda_1\partial_x+\lambda_2\exp\Big(\frac{\lambda_1(1-d)}{2}x+\frac{\lambda^2_1(1-d)^2-
 4\alpha }{4d}t
 \Big)(u+\beta)\partial_v,\end{equation}
 where  $\lambda_1$ and  $\lambda_2\not=0$ are arbitrary constants,
 of the RD system  \begin{equation}\begin{array}{l}
u_{xx}=u_t+(u+\beta)f(u),\\v_{xx}=dv_t+f(u)v+\alpha
v+g(u).\end{array}\end{equation}
  Finally, using the simple transformation   \begin{equation}\label{36}u\rightarrow u-\beta,\end{equation}
one sees that it is exactly case 6 of Table~1.

To complete the examination of case {\it (1)} we look for the
correctly-specified function $f(u)$, which satisfies (\ref{27*})
without the restriction $p^2=\beta q$. Indeed,  if one finds the
differential consequences of the second order (see equation for
$C^2$) then $C^2_{vx}=0, \ C^2_{vt}=0$  and  two algebraic equation
to find the function $f(u)$ are obtained:
\begin{equation}\nonumber\begin{array}{l}(q_tp^2-qp^2_t)f=((q_{xx}-dq_t)u+p^2_{xx}-dp^2_t)(q_tu+p^2_t)-\\
\qquad ((q_{xx}-dq_t)_tu+(p^2_{xx}-dp^2_t)_t)(qu+p^2)
 ,\\
 (q_xp^2-qp^2_x)f=((q_{xx}-dq_t)u+p^2_{xx}-dp^2_t)(q_xu+p^2_x)-\\
\qquad ((q_{xx}-dq_t)_xu+(p^2_{xx}-dp^2_t)_x)(qu+p^2),
\end{array}\end{equation}
 Thus,  $f(u)=\alpha_1+\alpha_2u+\alpha_3u^2$ provided  $p^2\neq\beta q$.
 Substituting this expression into (\ref{27*}) and
 making the standard routine, one  arrives at  case 8 of Table~1 if  $\alpha_3\neq0$  and case 9  if  $\alpha_3=0$.

Cases (2)--(6)  were  treated in the similar way and the results are
listed in Table~1. It should be noted that several  local
transformations ( (\ref{36}) is the simplest example)  were used to
reduce the number of cases and simplify  structures  of  the
relevant RD systems. These transformations can be presented  in the
general form (\ref{32}).

The sketch of  proof is now completed. $\blacksquare$

{\bf Table 1. $Q$-conditional symmetry operators  of the RD system
(\ref{3})  with \ $d\neq1.$ }
\begin{small}
\begin{center}
\begin{tabular}{|c|c|c|c|
} \hline &&& \\

  & $C^1(u,v)$& $C^2(u,v)$& $Q$   \\

\hline

&&&\\
1.&$uf(\omega)$&$u^kg(\omega)+u\big(f(\omega)+\alpha(1-d)\big)$&
 $\partial_t+\alpha u\p_u+\alpha\big((1-k)u+kv\big)\p_v, $
\\ &&$\omega=u^{-k}(v-u)$&$ \alpha\neq0, \ k\neq1$ \\
\hline &&&\\
2.&$uf(\omega)$&$u(g(\omega)+\alpha(1-d)\ln u+$&
 $\partial_t+\alpha u\p_u+\alpha(u+v)\p_v, \ \alpha \neq0$
\\ &&$f(\omega)\ln u), \ \omega=u\exp(-\frac{v}{u})$& \\
\hline &&&\\
3.&$ uf(\omega)$&$g(\omega)+u(f(\omega)+\alpha(1-d))$&
 $\partial_t+\alpha u\p_u+\alpha(u+1)\p_v, \ \alpha\neq0$
\\ &&$\omega=u\exp(u-v)$& \\
\hline &&&\\
4.&$f(\omega)$&$e^{u}g(\omega)-f(\omega)-\alpha(1-d)$&
 $\partial_t+\alpha\p_u+\alpha (u+v-1)\p_v, \ \alpha \neq0$
\\ &&$\omega=e^{- u}(u+v)$& \\
\hline &&&\\
5.&$f(\omega)$&$uf(\omega)+g(\omega)+(1-d)u$&
 $\partial_t+\p_u+u\p_v$
\\ &&$\omega=u^2-2v$& \\
\hline \end{tabular}
\end{center}
\end{small}

\begin{small}
\begin{center}

\begin{tabular}{|c|c|c|c|
} \hline  &&&\\
6.&$uf(u)$&$vf(u)+g(u)+\alpha v$&
 $\partial_t+\frac{2\lambda_1}{1-d}\p_x+qu\p_v,  \ \lambda_2\neq0$
\\  &&&$q=\lambda_2\exp\Big(\lambda_1x+\frac{\lambda^2_1-
 \alpha }{d}t
 \Big)$ \\
\hline &&&\\
 7.&$f(u)$&$(u+v)\Big(g(u)+\alpha\ln(u+v)\Big)-$&
 $\partial_t+\lambda \exp(-\frac{\alpha}{d} t)(u+v)\p_v,$
\\ &&$f(u)$&$\lambda\neq0$\\
\hline &&& \\

8.&$\alpha_1+\alpha_2u+$&$g(u)+uv$&
 $\partial_t+\frac{2\lambda_1}{1-d}\p_x+(qu+p^2)\p_v,$
\\
&$u^2$&&$q=\varphi_1(t)\exp(\lambda_1x), \ \varphi_1\neq0$
\\ &&&$p^2=\Big((\lambda^2_1+
 \alpha_2)\varphi_1-d\dot{\varphi}_1\Big)\exp(\lambda_1x)$ \\
\hline &&& \\
9.&$\alpha_1+\alpha_2u$&$g(u)+\alpha_3v$&
 $\partial_t+\frac{2\lambda_1}{1-d}\p_x+(qu+p^2)\p_v,$
\\
&&&$q=\lambda_2\exp\Big(\lambda_1x+\frac{\lambda^2_1+
 \alpha_2-\alpha_3 }{d}t
 \Big),$ \\ &&&$p^2_{xx}=dp^2_t+\alpha_3p^2-\alpha_1q, \ \lambda_2\neq0$ \\
\hline  &&& \\
 10.&$\alpha_1+\alpha_2u+$&$\alpha_3v+(\alpha_3-\alpha_2)u-$&
 $\partial_t+\big(\psi(x)\exp\big(\frac{\alpha_1\alpha_2}{\alpha_4(1-d)}t\big)-\frac{\alpha_1}{1-d}\big)
 (\p_u-$
\\
&$\alpha_4\ln(u+v)$&$\alpha_4\ln(u+v)$&$\p_v)+\frac{\alpha_1\alpha_2}{\alpha_4(1-d)}(u+v)\p_v,
\ \alpha_1\alpha_2\alpha_4\neq0$\\
  \hline &&& \\
 11.&$\alpha_2
u+v$&$\alpha_3v+\frac{1}{2}(1+d)\frac{v^2}{u}+$&
 $\partial_t+\varphi_2(t)(u\p_u+v\p_v)-\dot{\varphi}_2(t)u\p_v,
  $\\ &&$\alpha_1u\ln u +\alpha_4u$&$\alpha_2\dot{\varphi}_2(t)\neq0$ \\
\hline
 &&& \\ 12.&$\alpha_2 u$&$\alpha_3v+\alpha_4u+u^k$&
 $\partial_t+\lambda_2u\p_u+(\varphi_3(t)u+\lambda_2kv)\p_v,$ \\ &&&
  $\alpha_4\lambda_2\varphi_3\neq0, \ k\neq1$ \\
\hline &&& \\ 13.&$\alpha_1 u\ln u$&$\alpha_3v+\alpha_1v\ln u+$&
 $\partial_t+\frac{2\lambda_1}{1-d}\p_x+\lambda_2e^{-\alpha_1t}u\p_u+(qu+

 $ \\ &&$\alpha_2u^{\frac{1}{d}}$&$\frac{\lambda_2}{d}e^{-\alpha_1t}v)\p_v, \ q=\lambda_3\exp\Big(\lambda_1x+\frac{\lambda^2_1-
 \alpha_3 }{d}t+$
 \\&&&$\frac{d-1}{\alpha_1d}\lambda_2e^{-\alpha_1t}
 \Big), \quad \alpha_1\lambda_2\lambda_3\neq0$\\ \hline

&&& \\ 14.&$\alpha_1 u\ln u$&$\alpha_3v+\alpha_1v\ln u+\alpha_4u$&
 $\partial_t+\lambda_2e^{-\alpha_1t}u\p_u+
 \big(\varphi_4(t)u+(\frac{\lambda_2}{d}e^{-\alpha_1t}+
  $
 \\&&&$\lambda_3)v\big)\p_v, \ \alpha_1\lambda_2\varphi_4\neq0$\\
 \hline

&&& \\ 15.&$\alpha_1 u\ln u$&$\alpha_3v+\alpha_1v\ln u+$&
 $\partial_t+\lambda_2e^{-\alpha_1t}u\p_u+(\varphi_5(t)u+\frac{\lambda_2}{d}e^{-\alpha_1t}v)\p_v,
 $ \\ &&$\alpha_4u+\alpha_2u^{\frac{1}{d}}$&$\alpha_1\alpha_2\alpha_4\lambda_2\varphi_5\neq0$\\
\hline &&& \\ 16.&$\alpha_1 u\ln u$&$\alpha_3v+\alpha_1dv\ln u+$&
 $\partial_t+\lambda_2e^{-\alpha_1t}u\p_u+
 $ \\ &&$\alpha_1(1-d)u\ln u-\alpha_3u$&$(\alpha_1u+(\lambda_2e^{-\alpha_1t}-\alpha_1)v)\p_v, \ \alpha_1\lambda_2\neq0$\\

\hline &&& \\ 17.&$0$&$\alpha_3v+\ln u$&
 $\partial_t+\frac{2\lambda_1}{1-d}\p_x+\lambda_2u\p_u+(qu+p^2)\p_v, $
\\
&&&$q=\lambda_3\exp\Big(\lambda_1x+\frac{\lambda^2_1-\alpha_3+\lambda_2(1-d)
}{d}t
 \Big),$
 \\ &&&$p^2_{xx}=dp^2_t+\alpha_3p^2+\lambda_2, \ \lambda_2\lambda_3\neq0$ \\
\hline \end{tabular}
\end{center}
\end{small}

\begin{small}
\begin{center}

\begin{tabular}{|c|c|c|c|
} \hline

 &&& \\ 18.&$\alpha_2u$&$\alpha_3v+\ln u+\alpha_4u$&
 $\partial_t+\lambda_2u\p_u+(\varphi_6(t)u+p^2)\p_v, \ \alpha_4\lambda_2\varphi_6\neq0,$
 \\&&$$&$p^2_{xx}=dp^2_t+\alpha_3p^2+\lambda_2$\\
\hline &&& \\ 19.&$\alpha_2 u$&$\alpha_3v+u\ln u$&
 $\partial_t+\lambda_2u\p_u+(\varphi_7(t)u+\lambda_2v)\p_v, \  \lambda_2\varphi_7\neq0
$ \\
\hline &&& \\ 20.&\hskip1mm$\alpha_1+\alpha_2u+$&$uv+\alpha_3$&
 $\partial_t+\lambda\big(v+\varphi_8(t)u+\alpha_2\varphi_8(t)-
 d\dot{\varphi}_8(t)\big)\p_v,$
 \\  &$u^2$&& $\lambda\varphi_8\neq0$\\
\hline  &&& \\ 21.&$\alpha_1+\alpha_2u$&$\alpha_3v+u^2$&
 $\partial_t+\frac{2\lambda_1}{1-d}\p_x+p^1\p_u+(qu+p^2)\p_v,$
\\
&&&$q=\varphi_9(t)\exp(\lambda_1x), \ \varphi_9\neq0,$
 \\ &&&$p^2_{xx}=dp^2_t+\alpha_3p^2+(d-1)qp^1-\alpha_1q,$ \\ &&&$p^1=\frac{1}{2}
 \Big((\lambda^2_1+
 \alpha_2-\alpha_3)\varphi_9-d\dot{\varphi}_9\Big)\exp(\lambda_1x)$
 \\
\hline &&& \\ 22.&$0$&$\alpha_3v+e^{u}$&
 $\partial_t+\frac{2\lambda_1}{1-d}\p_x+\lambda_2\p_u+(qu+\lambda_2v+p^2)\p_v,$
 \\&&&$p^2_{xx}=dp^2_t+\alpha_3 p^2-\lambda_2(1-d)q,$\\
 &&&$q=\lambda_3\exp\Big(\lambda_1x+\frac{\lambda^2_1-
 \alpha_3}{d}t\Big), \ \lambda_3\neq0$\\
\hline &&& \\ 23.&$\alpha_1$&$\alpha_3v+\alpha_4u+$&
 $\partial_t+\lambda_2\p_u+(\varphi_{10}(t)u+\lambda_2v+p^2)\p_v, \  \alpha_4\lambda_2\varphi_{10}\neq0,$
 \\&&$e^{u}$&$p^2_{xx}=dp^2_t+\alpha_3 p^2-(\lambda_2(1-d)+\alpha_1)q+\alpha_4\lambda_2$\\
\hline &&& \\ 24.& 0&$\alpha_1(u+v)\times$&
 $\partial_t+\frac{\alpha_2}{d-1}(\p_u-\p_v)+\lambda\exp(-\frac{\alpha_1}{d}t)(u+v)\p_v, $
\\&&$\ln(u+v)+\alpha_2$&$\alpha_1\alpha_2\lambda\neq0$ \\
\hline
 &&& \\ 25. & 0&$u^2$&
 $\quad 2t\partial_t+x\p_x+\frac{2t-x^2}{4\sqrt{t^5}}\exp(-\frac{x^2}{4t})\p_u
 +\big(\frac{1}{\sqrt{t}}\exp(-\frac{x^2}{4t})u+$
\\ &&&$2v+p^2\big)\p_v, \ d=3, \  p^2_{xx}=3p^2_t+\frac{2t-x^2}{4t^4}\exp(-\frac{x^2}{2t}), $\\
\hline  &&& \\
26.&  0&$ u^5   $&
 $t^2\partial_t+tx\p_x-\frac{x^2+2t}{4}u\p_u+\big(\lambda\exp(-\frac{x^2}{t})u
 -  $
\\ &&&$\frac{5x^2+2t}{4}v\big)\p_v, \ d=5,  \ \lambda\neq0$\\
\hline

\end{tabular}
\end{center}
\end{small}

 In Table~1, the functions  $\psi(x)$,
$\varphi_1(t)$, $\varphi_2(t)$, $\varphi_4(t)$,  $\varphi_5(t)$,
$\varphi_8(t)$, and $\varphi_9(t)$ are the general solutions of the
equations

\[ \psi''-\big(\frac{\alpha_1\alpha_2}{\alpha_4(1-d)}+\alpha_2\big)\psi=0,\]
\[ d\ddot{\varphi}_1-(2\lambda^2_1+
 \alpha_2)\dot{\varphi}_1+\frac{\lambda^4_1+\alpha_2\lambda^2_1+
 \alpha_1}{d}\varphi_1=0,\]
\[ d\ddot{\varphi}_2-(\alpha_2-\alpha_3+(1-d)\varphi_2)\dot{\varphi}_2-\alpha_1\varphi_2=0,\]
\[ d\dot{\varphi}_4+(\alpha_3+\lambda_2(d-1)e^{-\alpha_1t})\varphi_4-\alpha_4\big(\lambda_3+
\lambda_2\frac{1-d}{d}e^{-\alpha_1t}\big)=0,\]
\[
d\dot{\varphi}_5+(\alpha_3+\lambda_2(d-1)e^{-\alpha_1t})\varphi_5-\alpha_4
\lambda_2\frac{1-d}{d}e^{-\alpha_1t}=0,\]
\[ d\ddot{\varphi}_8-\alpha_2\dot{\varphi}_8+
\frac{\alpha_1}{d}\varphi_8+\frac{\alpha_3}{d}=0,\] and
\[ d\ddot{\varphi}_9-\big(\lambda^2_1(1+d)+\alpha_2(1-d)-
\alpha_3\big)\dot{\varphi}_9+\big(\lambda^4_1-\alpha_3\lambda^2_1-
 \alpha_2(\alpha_2-\alpha_3)\big)\varphi_9=0,\]
 respectively. The functions
\[\varphi_3(t)=
\begin{cases} \lambda_3\exp\Big(\frac{
 \alpha_2-\alpha_3+\lambda_2(1-d)}{d}t
 \Big)+\frac{\alpha_4\lambda_2(1-k)}{\alpha_2-\alpha_3+\lambda_2(1-d)},
  & \texttt{if} \quad  \alpha_2\neq \alpha_3-\lambda_2(1-d), \\-\frac{\alpha_4\lambda_2(1-k)}{d}t+\lambda_3,
  & \emph{if} \quad \alpha_2=\alpha_3-\lambda_2(1-d); \end{cases} \]
\[\varphi_{10}(t)=
\begin{cases} \lambda_3\exp(-\frac{
 \alpha_3}{d}t)+\frac{\alpha_4\lambda_2}{\alpha_3},
  &  \texttt{if} \,\alpha_3\neq0, \\ \frac{\alpha_4\lambda_2}{d}t,
  & \texttt{if} \, \alpha_3=0. \end{cases} \]
Finally, the function  $\varphi_6(t)= \varphi_3(t)$  at  $k=0, $
while $\varphi_7(t)=\varphi_3(t)$  at $k=0$  and $\alpha_4=1.$
 Hereafter the upper  dot  index denotes differentiation with respect to the
 variable $t$.

\section{\bf Form-preserving transformations of the RD systems}

A natural question is: Can we claim that 26 systems  listed in
Table~1 are inequivalent up to any local substitutions (not only of
the form (\ref{32})!)? It turns out that the answer is positive. To
present the rigorous proof  of this, we used   the notion of the set
of form-preserving point transformations introduced  in
\cite{kingston-91} and  now extensively used for Lie symmetry
classification problems (see, e.g.
\cite{ch-se-ra-08,van-pop-sop-09}). Note that finding these
transformations for systems of PDEs is  a difficult problem because
of technical problems occurring in computations and there is no many
results for systems (paper  \cite{ch-myr-2010} is one of the first
presenting  an explicit  result for a class of systems).

 The form-preserving  transformations present the
most general  and correctly-specified form
 of local  substitutions, which can map {\it some  equations}  from a given class to other  those  belonging to the same class.
 They  contain as  particular cases the well-known equivalence transformations
 and discrete transformations, which maps  {\it each  equation}  from the  class to another one from this class,    used in the well-known  Ovsiannikov method of Lie symmetry classification.
 Here we construct  such transformations  with the aim to show that Table~1 cannot be shortened.

\bt An arbitrary  RD system of the form (\ref{3}) with $d\neq 1$ can
be reduced to another   system of the same form
\begin{equation}\label{2-7}\begin{array}{l} w_{yy}=w_\tau+F^1(w,z),
\\z_{yy}=\lambda z_\tau+F^2(w,z),\end{array}\end{equation} by the  non-degenerate local 
transformation
\begin{eqnarray}&&\label{2-11}
\tau=a(t,x,u,v), \ y=b(t,x,u,v), \\&& \label{2-12}
w=\varphi(t,x,u,v), \ z=\psi(t,x,u,v),\end{eqnarray} if and only if
the smooth functions $a, \ b, \ \varphi$ and $\psi$ satisfy one of two sets conditions listed below.
\begin{equation}\label{2-8}\begin{array}{l}(I)
\hskip5mm a=\alpha(t), \ b=\beta(t)x+\gamma(t), \ \alpha \beta \neq0,\\
\hskip10.7mm \varphi=f(t)\exp\Big(-\frac{1}{4\beta}(\dot{\beta}x^2+
2\dot{\gamma}x)\Big)u+P(t,x),\ f\neq0,\\
\hskip10.7mm \psi=g(t)\exp\Big(-\frac{d}{4\beta}(\dot{\beta}x^2+
2\dot{\gamma}x)\Big)v+Q(t,x),\ g\neq0,
\end{array}\end{equation}  where the functions $\alpha(t),  \beta(t), f(t),
g(t), \gamma(t), P(t,x),$ and $Q(t,x)$ are such that  the
equalities
\begin{eqnarray}&&\label{2-9}\dot{\alpha}=\beta^2, \quad \lambda=d,
\\&&\label{2-10}
\beta^2F^1(\varphi,\psi)=\varphi_uC^1(u,v)+
\varphi_{xx}-\varphi_t-
2\frac{\varphi_x}{\varphi_u}\varphi_{xu},\
\\&&\label{2-13}\beta^2F^2(\varphi,\psi)=\psi_vC^2(u,v)+\psi_{xx}
-d\psi_t-2\frac{\psi_x}{\psi_v}\psi_{xv} \end{eqnarray} take place; \be\label{d140}\ba{l} (II) \hskip3mm a=\alpha(t), \ b=\beta(t)x+\gamma(t), \ \alpha \beta \neq0,\\
\hskip10.7mm \varphi=f(t)\exp\Big(-\frac{d}{4\beta}(\dot{\beta}x^2+
2\dot{\gamma}x)\Big)v+P(t,x), \ f\neq0,\\
\hskip10.7mm \psi=g(t)\exp\Big(-\frac{1}{4\beta}(\dot{\beta}x^2+
2\dot{\gamma}x)\Big)u+Q(t,x), \ g\neq0, \ea\ee where the functions
$\alpha(t), \  \beta(t), \ \gamma(t), \ f(t), \ g(t),
 \ P(t,x)$ and $Q(t,x)$ are such that  the
equalities
\begin{eqnarray}
& &\dot{\alpha}=\lambda\beta^2, \label{d141}\quad
\lambda=\frac{1}{d},  \\
& &\beta^2F^1(\varphi,\psi)=\varphi_vC^2(u,v)+
\varphi_{xx}-d\varphi_t-2\frac{\varphi_x}{\varphi_v}\varphi_{xv},\label{d142} \\
& &\beta^2F^2(\varphi,\psi)=\psi_uC^1(u,v)+\psi_{xx}
-\psi_t-2\frac{\psi_x}{\psi_u}\psi_{xu}. \label{d143}
\end{eqnarray} take place.
 \et

 {\bf Proof.} First of the all we note
that  each non-degenerate transformation
  (\ref{2-11}) -- (\ref{2-12})  should satisfy the condition
\begin{equation}\label{2-14} \Delta_1=\begin{vmatrix} a_x & a_t & a_u & a_v\\
b_x & b_t & b_u & b_v \\ \varphi_x & \varphi_t & \varphi_u & \varphi_v \\
\psi_x & \psi_t & \psi_u & \psi_v
\end{vmatrix}\neq0, \end{equation}
which is  used to prove the theorem.

Let us choose an arbitrary  RD system of the form (\ref{3}). The
main idea of the proof is based on substituting the expressions for
$u_{xx}, \ v_{xx}, \ u_t, \ v_t$   using the formulae  (\ref{2-11})
and (\ref{2-12}) into  this system and on analysis conditions when
the system obtained is equivalent to system (\ref{2-7}).
 The expressions for the first-order derivatives have the form

\begin{equation}\nonumber\begin{array}{l} \medskip u_x=\frac{\begin{vmatrix}
\varphi_x-a_xw_\tau-b_xw_y & a_vw_\tau+b_vw_y-\varphi_v \\
\psi_x-a_xz_\tau-b_xz_y & a_vz_\tau+b_vz_y-\psi_v
\end{vmatrix}}{\begin{vmatrix} a_uw_\tau+b_uw_y-\varphi_u & a_vw_\tau+b_vw_y-\varphi_v \\
a_uz_\tau+b_uz_y-\psi_u &  a_vz_\tau+b_vz_y-\psi_v
\end{vmatrix}}, \\
u_t=\frac{\begin{vmatrix}
\varphi_t-a_tw_\tau-b_tw_y & a_vw_\tau+b_vw_y-\varphi_v \\
\psi_t-a_tz_\tau-b_tz_y & a_vz_\tau+b_vz_y-\psi_v
\end{vmatrix}}{\begin{vmatrix} a_uw_\tau+b_uw_y-\varphi_u & a_vw_\tau+b_vw_y-\varphi_v \\
a_uz_\tau+b_uz_y-\psi_u
 &  a_vz_\tau+b_vz_y-\psi_v
\end{vmatrix}}. \end{array}\end{equation}
The expressions for the second-order derivatives  are very
cumbersome, however, it can be noted that  they contain the
derivative $w_{\tau\tau}$ and $w_{\tau y}$. Because   $\tau$ is  a
new time-variable we conclude that the coefficient next to
$w_{\tau\tau}$ and $w_{\tau y}$ must vanish otherwise system
(\ref{2-7}) are not obtainable. These coefficients  vanish if and
only if  the equalities take place: \begin{equation}\label{2-19}
a_x=a_u=a_v=b_u=b_v=0 \ \Rightarrow \ a=\alpha(t), \ b=b(t,x).
\end{equation} Moreover, taking into account  (\ref{2-14}),  the
restrictions \be\nonumber \dot{\alpha} \
b_x\neq0, \qquad \Delta_2=\begin{vmatrix} \varphi_u & \varphi_v \\
\psi_u & \psi_v
\end{vmatrix}\neq0 \ee are  also obtained.

Having the set of  equalities (\ref{2-19}),  the expressions for
$u_{xx}$ and $u_t$   can be essentially  simplified, namely:
\begin{equation}\label{2-20}\begin{array}{l} \medskip
u_{xx}=\frac{\psi_vb^2_x}{\Delta_2}w_{yy}-\frac{\varphi_vb^2_x}{\Delta_2}z_{yy}+
\frac{(\psi_vb_x)_x\Delta_2-(\Delta_2)_x\psi_vb_x}{\Delta^2_2}w_{y}-
\frac{(\varphi_vb_x)_x\Delta_2-(\Delta_2)_x\varphi_vb_x}{\Delta^2_2}z_{y}+\\
 \qquad \
\frac{(\psi_x\varphi_v-\psi_v\varphi_x)_x\Delta_2-
(\Delta_2)_x(\psi_x\varphi_v-\psi_v\varphi_x)}{\Delta^2_2}, \\
u_t=\frac{1}{\Delta_2}\big(\psi_v(\dot{\alpha}w_\tau+b_tw_y-\varphi_t)-
\varphi_v(\dot{\alpha}z_\tau+b_tz_y-\psi_t)\big).
\end{array}\end{equation} Substituting   (\ref{2-20}) into the first
equation of  (\ref{3}). Omitting the full  expression of the
equation obtained, we note that one  contains the terms
 \begin{equation}\nonumber(i) \
-\frac{\varphi_vb^2_x}{\Delta_2}(z_{yy}-\frac{\dot{\alpha}}{b^2_x}z_\tau),
\qquad (ii) \
\frac{\psi_vb^2_x}{\Delta_2}(w_{yy}-\frac{\dot{\alpha}}{b^2_x}w_\tau),\end{equation}
while other terms don't depend on  $z_{yy},
w_{yy}, z_\tau, $ and $ w_\tau.$ \\

 Now there is two possibilities. If the first equation of (\ref{3}) is transformed into the first one of  (\ref{2-7}) then we immediately obtain
 \begin{equation}\label{2-21} \dot{\alpha}=b^2_x, \ \varphi_v=0.\end{equation}
If the first equation of (\ref{3}) is transformed into  the second
one then  the conditions \begin{equation}\label{2-25}
\dot{\alpha}=\lambda b^2_x, \ \psi_v=0\end{equation}  must be
satisfied.

Let us consider  conditions  (\ref{2-21}).  Taking into account
(\ref{2-14}) and $\varphi_v=0$,  the restriction
$\varphi_u\psi_v\neq 0$ springs up.

On the other hand,
$b(t,x)=\beta(t)x+\gamma(t),$  follows from (\ref{2-21})  where
$\beta$  and   $\gamma$  are arbitrary smooth functions. Thus, the
first equation from (\ref{2-9}) is derived.

Substituting  (\ref{2-21}) into  expressions for   $u_{xx}$ and
$u_t$ (see formulae (\ref{2-20})), one obtains
\begin{equation}\label{2-23}\begin{array}{l}u_{xx}=\frac{\beta^2}{\varphi_u}w_{yy}-2\frac{\beta\varphi_{xu}}{\varphi^2_u}w_y
+\frac{2\varphi_x\varphi_{xu}-\varphi_u\varphi_{xx}}{\varphi^2_u}-
\frac{\varphi_{uu}}{\varphi^3_u}(\beta w_y-\varphi_x)^2, \\
u_t=\frac{1}{\varphi_u}(\dot{\alpha}w_\tau+(\dot{\beta}x+\dot{\gamma})w_y-\varphi_t).\end{array}\end{equation}
Since the first equation of system  (\ref{2-7})  doesn't contain the
terms $w_y$ and $w_y^2$, we should vanish the relevant coefficient,
namely: \[\begin{array}{l}
2\frac{\beta\varphi_{xu}}{\varphi^2_u}+\frac{\dot{\beta}x+
\dot{\gamma}}{\varphi_u}=0,\\  \varphi_{uu}=0. \end{array}\] The
general solution of this system can be easily constructed so that
obtains \begin{equation}\label{2-24}
\varphi=f(t)\exp\Big(-\frac{1}{4\beta}(\dot{\beta}x^2+
2\dot{\gamma}x)\Big)u+P(t,x),\end{equation} where  $f(t)\neq0$  and
$P(t,x)$ are arbitrary functions at the moment.
Thus, the first, second and third  equations from  (\ref{2-8}) are
derived.
 Moreover, substituting    (\ref{2-23})  and  (\ref{2-24})  into the first equation of system (\ref{3}),  we arrive at the equation
\begin{equation}\label{2-26}
w_{yy}=w_\tau+\frac{\varphi_u}{\beta^2}\big(C^1(u,v)-\frac{\varphi_t}{\varphi_u}-\frac{2\varphi_x\varphi_{xu}-
\varphi_u\varphi_{xx}}{\varphi^2_u}\big).\end{equation} Now one
realizes that (\ref{2-26})  coincides with the first  equation of
system (\ref{2-7}) iff   condition (\ref{2-10})  takes place.

The analogous routine involving the second equation of system
(\ref{3}) leads to the condition
$\dot{\alpha}=\frac{\lambda}{d}\beta^2 \Rightarrow \lambda=d$ (see
(\ref{2-21})),
 the function $\psi$  of the form  (\ref{2-8})  and equation  (\ref{2-13}).

Analogous examination of conditions (\ref{2-25})   leads to  transformations
(\ref{d140}) and to  equalities (\ref{d141})--(\ref{d143}).

The proof is now completed. $\blacksquare$

\medskip

\textbf{ Consequence 1.} The set of transformations (\ref{32})
arising in theorem 1 is a subset of form-preserving transformations
(\ref{2-8}).

\textbf{ Consequence 2.} If  the nonlinear   RD system of the form
(\ref{1})
   is transformed  to another one from this class, say, to the system
     \begin{equation}\nonumber\begin{array}{l}
u^*_{t^*}=d^*_1u^*_{x^*x^*}+F^*(u^*,v^*), \\
v^*_{t^*}=d^*_2v^*_{x^*x^*}+G^*(u^*,v^*)\end{array}\end{equation}
    by a local  substitution then they have the proportional diffusivities.
     Moreover, there  are two    linear combinations for
 the reaction terms $ F$ and $ F^*$,  and  for  $G $ and $ G^*$  resulting $\alpha_1u +\alpha_2$  and  $\alpha_3v +\alpha_4$, respectively (here $\alpha_k, k=1,\dots,4$ are correctly-specified constants).

   \medskip

   Roughly speaking,  consequence 2  says that the locally-equivalent RD systems  have the same structure up to additive terms $\alpha_1u +\alpha_2$  and  $\alpha_3v +\alpha_4$. At the first sight, there are some systems in Table~1 satisfying this consequence, for example in cases 17 and 18. However, according to consequence 2,  the term $ \alpha_4u$  arising in the second equation of  the RD system (see case 18) cannot be removed by any local substitution.
   We have carefully checked all cases listed in Table~1 and concluded that there are no any  locally-equivalent systems therein.

    Thus,
   we   have shown   that the list of RD systems  presented in  Table~1 cannot be reduced (shortened)  by any local substitution.

\section{\bf  New exact solutions and their possible interpretation}

It is well-known that using  the known  $Q$-conditional symmetry
(non-classical symmetry), one reduces  the given system of PDEs  to
a system of ODEs  via  the same procedure as for classical Lie
symmetries.  Since  each   $Q$-conditional symmetry of the first
type is automatically  one of the  second type, i.e.,
non-classical symmetry, we  apply  this  procedure  for finding
exact solutions. Thus, to construct an ansatz corresponding to the
given operator $Q$, the system of the linear first-order PDEs
\begin{equation}\label{45}Q\,(u)=0,\quad  Q\,(v)=0\end{equation}
 should be solved. Substituting the ansatz obtained
into the RD system  with correctly-specified coefficients, one
obtains the reduced system of ODEs.

Let us construct  exact solutions of  the non-linear  RD system
listed in the case 1 of Table~1, when the system and  the
corresponding symmetry operator have the form

\begin{equation}\label{43}\begin{array}{l}u_t=u_{xx}-uf(\omega),\\
dv_t=v_{xx}-u^kg(\omega)-u(f(\omega)+\alpha(1-d)), \,
\omega=u^{-k}(v-u)\end{array}\end{equation} and
 \begin{equation}\label{44}Q=\partial_t+\alpha
u\partial_u+\alpha\big((1-k)u+kv\big)\partial_v,\end{equation}
 In this case system  (\ref{45}) takes the form   \begin{equation}\label{46}\begin{array}{l}u_t=\alpha u, \\
v_t=\alpha(1-k)u+\alpha
kv \end{array}\end{equation}  and its general solution  produces the ansatz (the functions  $u$ and  $v$  depend on two variables $t$ and $x$):\begin{equation}\label{47}\begin{array}{l}u=\varphi(x)e^{\alpha t},\\
v=\psi(x)e^{k\alpha t}+\varphi(x)e^{\alpha
t},\end{array}\end{equation} where $\varphi(x)$ and  $\psi(x)$  are
new unknown functions. Substituting ansatz (\ref{47}) into
(\ref{43}), one obtains so called reduced system of ODEs
\begin{equation}\label{48}\begin{array}{l}\varphi''=\varphi\big(\alpha+f(\omega)\big), \\
\psi''=\varphi^kg(\omega)+\alpha kd\psi, \,
\omega=\psi\varphi^{-k}.\end{array}\end{equation}

Because system (\ref{48}) is non-linear (excepting, of course, some
special cases)  it can be integrated only for the
correctly-specified functions
 $f$  and $g$. We specify $f$  and $g$ in a such way, when the RD system in question will be still non-linear
  (otherwise the result will be rather trivial). Thus, setting
$f(\omega)=\gamma\omega^{\frac{1}{k}}-\alpha, \
g(\omega)=\beta\omega,$ ( $\beta$ and $ \gamma$  are arbitrary
non-zero constants, the RD system  takes the form
\begin{equation}\label{49}\begin{array}{l}u_t=u_{xx}-\gamma(v-u)^{\frac{1}{k}}+\alpha u,\\
dv_t=v_{xx}-\gamma(v-u)^{\frac{1}{k}}-\beta v+(\beta+\alpha d)u,
\end{array}\end{equation} while the corresponding reduced system is
\begin{equation}\label{50}\begin{array}{l}\varphi''=\gamma\psi^{\frac{1}{k}}, \\
\psi''=(\beta+\alpha kd)\psi.\end{array}\end{equation}  The general
solution of (\ref{50}) can be easily constructed:
\begin{equation}\label{51}\varphi(x)=\gamma\int\Big(\int\psi^{\frac{1}{k}}(x)dx\Big)dx+c_3x+c_4,\end{equation}
\begin{equation}\label{51*} \psi(x)=
\begin{cases} c_1\exp(\mu x)+c_2\exp(-\mu x),  & \text{if} \ \mu^2=\beta+\alpha
kd>0, \\  c_1\cos(\nu x)+c_2\sin(\nu x),  &  \text{if} \
\nu^2=-(\beta+\alpha kd)>0,  \\  c_1x+c_2,  & \text{if} \
\beta+\alpha kd=0. \end{cases} \end{equation} Thus,  substituting
(\ref{51}) and (\ref{51*}) into  (\ref{47}), the 4-parameter family
of solutions for the non-linear RD system  (\ref{49}) is
constructed.

Hereafter we highlight the solutions satisfying  the zero Neumann
boundary conditions, which widely arise in biologically motivated
boundary-value problems. Hence, setting  $c_3=c_4=0, \
k=\frac{1}{3}, \ \psi=c_1\cos(\nu x)$, one obtains the solution
\begin{equation}\nonumber\begin{array}{l}u=-\gamma\frac{c^3_1}{9\nu^2}(\cos^2(\nu
x)+6)\cos(\nu x)e^{\alpha t}, \\ v=c_1\cos(\nu
x)e^{\frac{1}{3}\alpha t}+u.\end{array}\end{equation} It can be
noted that this solution satisfies
 the zero Neumann  boundary conditions
 \begin{equation}\nonumber u_x|_{x=0}=0, \ v_x|_{x=0}=0,  \
u_x|_{x=j\frac{\pi}{\nu}}=0, v_x|_{x=j\frac{\pi}{\nu}}=0
\end{equation} on the interval $[0, j\frac{\pi}{\nu}]$, where $ j
\in \mathbb{N}$.

Let us set  $f(\omega)=-(a_1+b\omega)$,
$g(\omega)=(\alpha(1-d)-a_1)\omega$ (hereafter $\alpha,  a_1$ and
$b$  are arbitrary non-zero constants)  in  (\ref{43}), hence, it
takes the form
\begin{equation}\label{52}\begin{array}{l}u_t=u_{xx}+a_1u-bu^{2-k}+bvu^{1-k},\\
dv_t=v_{xx}-a_2v-bu^{2-k}+bvu^{1-k}, \end{array}\end{equation} where $a_2=\alpha(1-d)-a_1$. 
The corresponding reduced system of ODEs is
\begin{equation}\label{54}\begin{array}{l}\varphi''+b\psi\varphi^{1-k}+(a_1-\alpha)\varphi=0, \\
\psi''=(\alpha kd+a_2)\psi.\end{array}\end{equation} Nevertheless we
have not constructed  the general solution of system (\ref{54}),
  its  particular solution was found by setting $\psi=-\delta, \, \delta\not=0.$
  In this case, the first-order ODE
\begin{equation}\label{56}\varphi'=\pm\sqrt{(\alpha-a_1)\varphi^2+\frac{2b\delta}{2-k}\varphi^{2-k}+c_1},\,
\ \alpha=\frac{a_1}{d(k-1)+1}, \ k\neq2 \end{equation}  for  the
function $\varphi$ is obtained ( the value $k=2$ is special and
leads to ODE
$\varphi'=\pm\sqrt{(\alpha-a_1)\varphi^2+2b\delta\ln\varphi+c_1}$).

\begin{figure}[t]\label{fig-1}
\centerline{\includegraphics[width=6.4cm]{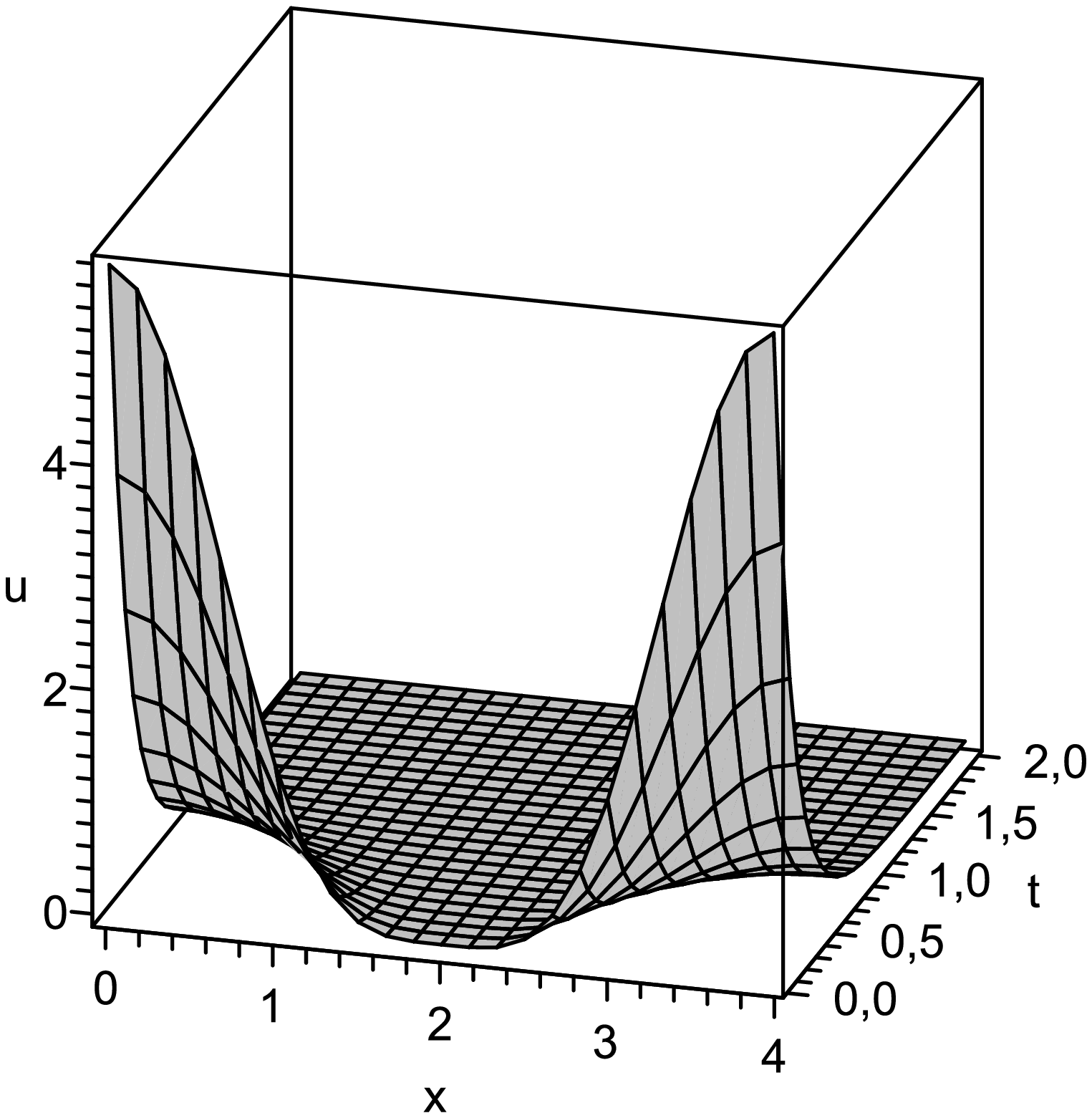},
\includegraphics[width=6.4cm]{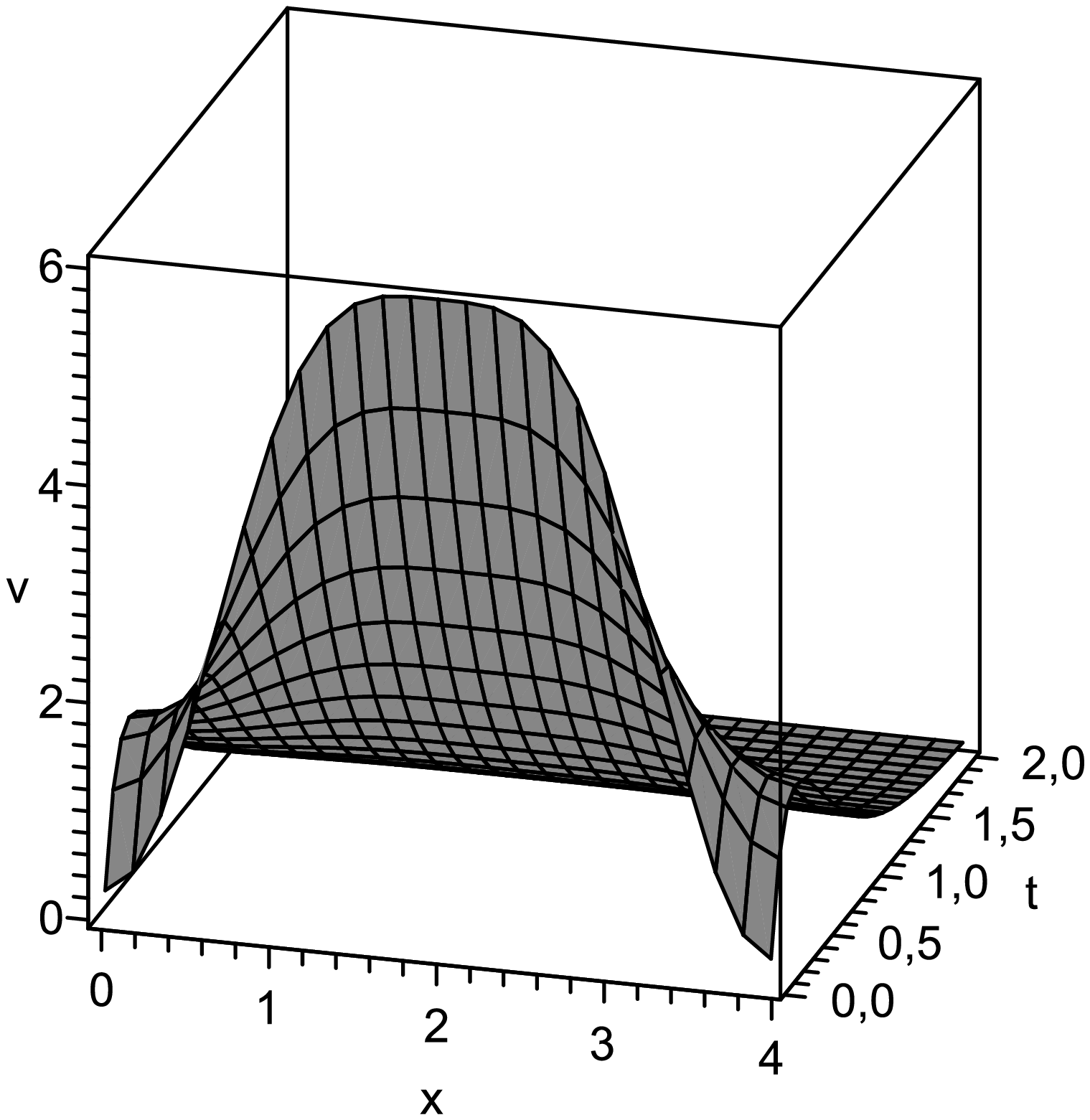}}
 \caption{ Exact
solution  (\ref{59})  with  $ k=0.5, \ \delta=6, \ d=4, \ a_1=5, \
b=3.$} \label{fig-1}
\end{figure}

If  $c_1\neq0$  then the general solution of  (\ref{56}) can be
expressed via hypergeometric  functions. Here we present the
solution for (\ref{56})  with  $c_1=0, \ k\neq0$:

\begin{equation}\label{57}\varphi(x)=
\begin{cases} \Big(\beta(\tan^2\big(\frac{k\sqrt{a_1-\alpha}}{2}(x\pm c_2)\big)+
1)\Big)^{-\frac{1}{k}},
  & a_1>\alpha, \\  \Big(-\beta(\tanh^2\big(\frac{k\sqrt{\alpha-a_1}}{2}(x\pm c_2)\big)-1)\Big)^{-\frac{1}{k}},
  & a_1<\alpha, \end{cases} \end{equation} ( $\beta=\frac{(a_1-\alpha)(2-k)}{2b\delta}$),
    which seems to be the most  interesting. Note that another   arbitrary constant can be removed by the trivial substitution
   $x\pm c_2\rightarrow x.$

Now we rewrite system  (\ref{52})  setting $v\rightarrow -v$  with
the aim  to obtain a biologically motivated model. So the system
takes form

\begin{equation}\label{58}\begin{array}{l}u_t=u_{xx}+u(a_1-bu^{1-k})-bvu^{1-k},\\
dv_t=v_{xx}+v(-a_2+bu^{1-k})+bu^{2-k},\end{array}\end{equation}
where all coefficients (excepting $k$) should be positive.
 (\ref{58})  can be treated as a  prey-predator model for the population dynamics.
In fact, the species $u$ is prey  and described by the first
equation.  Its population decreases proportionally to the predator
density $v$. The natural birth-dead rule for the prey   is
$u(a_1-bu^{1-k})$ and can be treated as a generalization of the
standard logistic rule  $u(a_1-bu)$ (see, e.g., \cite{mur2}). The
similar arguments are also valid  for the second equation. The model
should involve also  the zero Neumann boundary conditions (zero-flux
on the boundaries), which indicate that both species cannot
widespread over the globe  but occupy   a bounded domain.

Using  (\ref{47}) with  $v\rightarrow -v$, $\psi=-\delta$  and
(\ref{57}) with  $a_1>\alpha$   we construct the exact  solution
\begin{equation}\label{59}\begin{array}{l}u=\Big(\beta(\tan^2(\frac{k\sqrt{a_1-
\alpha}}{2}x)+1)\Big)^{-\frac{1}{k}}e^{\alpha t},\\ v=\delta
e^{\alpha kt}-\Big(\beta(\tan^2(\frac{k\sqrt{a_1-
\alpha}}{2}x)+1)\Big)^{-\frac{1}{k}}e^{\alpha
t}\end{array}\end{equation} of (\ref{58}). It turns out that
solution can  describe interaction between prey and predator on the
space  interval $[0, l]$, (here $l=\frac{2\pi j}{k\sqrt{a_1-
\alpha}}, \  j \in \mathbb{N}$) provided \begin{equation}\label{60}\
0<k<1-\frac{1}{d}, \
0<\delta\leq\Big(\frac{(2-k)(a_1-\alpha)}{2b}\Big)^{\frac{1}{1-k}},
\ \alpha=\frac{a_1}{d(k-1)+1}<1. \end{equation} One easily checks
that solution (\ref{59}) is non-negative, bounded in the domain
$\Omega=\{ (t,x) \in (0,+ \infty ) \times (0,l)\} $ and  satisfy the
given  zero Neumann boundary conditions, i.e.
\begin{equation}\nonumber u_x|_{x=0}=0, \ v_x|_{x=0}=0, \ u_x|_{x=l}=0, \
v_x|_{x=l}=0. \end{equation} As  example we present   this solution
(\ref{59}) with the  parameters satisfying the restrictions
(\ref{60}) in Fig.~\ref{fig-1}. This  solution can describe such
type of the interaction between the  species $u$ and $v$ when both
of them eventually die, i.e. $(u, v) \to (0,0)$ if $t \to +\infty.$

\section{\bf Conclusions}

In this paper,  $Q$-conditional symmetries for
 the class of RD  systems  (\ref {3}) (that  is  equivalent to the class of systems    (\ref {1}))  and their application for finding exact solutions are studied.
Following the recent paper \cite{ch-2010},    the notion  of
$Q$-conditional symmetry of the first type was used for these
purposes. The main result is presented in theorem 1 giving the
exhaustive list of RD systems of the form   (\ref {1}) with $d_1
\not=d_2 $ (the case $d_1 =d_2 $  should be analyzed separately),
which admit such symmetry. It turns out that there are  exactly 26
locally-inequivalent  RD systems  admitting the $Q$-conditional
symmetry operators of the first type of the form   (\ref{2-2}) with
$\xi^0 \not=0$
(the case $\xi^0 =0$ should be analyzed separately). To show  local
non-equivalence of the systems listed in Table~1, we proved  theorem
2  describing  the set of form-preserving point transformations for
the class of RD  systems  (\ref {3}). Note that all the operators
found are inequivalent to the Lie symmetry operators presented in
\cite{ch-king, ch-king2} because the necessary and sufficient
condition, which guarantees this property, was used.

The    $Q$-conditional operator listed in case 1 of Table~1 was used
to construct the non-Lie ansatz and to reduce two nonlinear  RD
systems  to  the corresponding ODE systems. Solving these ODE
systems, the  two-parameter  families  of exact solutions     were
explicitly constructed
 for the RD systems in question.
 Moreover,    application of  the exact solutions
   for solving  the prey-predator  system   (\ref {58})
   was presented. It  turns out  that the relevant
    boundary value problem  with the zero Neumann   conditions
    can be exactly  solved and the solution  can describe   the densities of two interacting species.

    The work is in progress to construct conditional  symmetries for   {\it multicomponent}  RD  systems. In particular case, a wide list of the   $Q$-conditional symmetries  of the first type for the three-component diffusive Lotka-Volterra system  is presented in  \cite{ch-dav-2013}.

  Finally, we point out  that this paper is a natural continuation of the recent paper
    \cite{ch-dav-2012}, where  RD  systems with non-constant diffusivities were examined.

\section{Acknowledgment}   R.Ch. thanks the Organizing Committee of
the 7th  Workshop `Algebra, Geometry, and Mathematical Physics'
(Mulhouse,  24-26 October 2011) for the  financial support.


\begin{thebibliography}{99.}

\footnotesize

\bibitem {ames} Ames, W.F.: Nonlinear Partial Differential
Equations in Engineering. Academic, New York (1972)

\bibitem {mur2} Murray, J.D.: Mathematical Biology. Springer, Berlin (1989)

\bibitem {mur2003} Murray, J.D.: Mathematical Biology  II: Spatial
 Models and Biomedical Applications. Springer, Berlin (2003)

\bibitem {aris} Aris, R.: The Mathematical Theory of  Diffusion and
 Reaction  in Permeable Catalysts.  Clarendon, Oxford (1975)

\bibitem {okubo} Okubo, A., Levin, S.A.: Diffusion and Ecological Problems.
Modern Perspectives, 2nd edn. Springer, Berlin (2001)

\bibitem {zulehner-ames} Zulehner, W., Ames, W.F.:  Group analysis of
a semilinear vector diffusion equation. Nonlinear Analysis
\textbf{7}, 945--69 (1983)

\bibitem {ch-king}  Cherniha, R., King, J.R.:
 Lie    Symmetries of Nonlinear  Multidimensional
Reaction-Diffusion Systems: I.  J. Phys. A: Math. Gen. \textbf{33},
267--82, 7839--41 (2000)

\bibitem {ch-king2} Cherniha, R., King, J.R.:
 Lie    Symmetries of Nonlinear  Multidimensional
Reaction-Diffusion Systems: II. J. Phys. A: Math. Gen. \textbf{36},
405--25 (2003)


\bibitem {niki-05} Nikitin, A.G.:  Group classification of systems of non-linear reaction-diffusion
equations.  Ukrainian Math. Bull. \textbf{2}, 153-204 (2005)

\bibitem{bl-c} Bluman, G.W., Cole, J.D.:
 The general similarity solution of the heat equation
J.~Math. Mech. {\bf 18}, 1025-42 (1969)

\bibitem {Fush93} Fushchych, W.I., Shtelen, W.M., Serov, M.I.:
Symmetry Analysis and Exact Solutions of Equations of Nonlinear
Mathematical Physics. Kluwer, Dordrecht (1993)

\bibitem{ch-he-2004} Cherniha, R. Henkel, M.:
On nonlinear partial differential  equations with an
infinite-dimensional conditional symmetry. J. Math. Anal. Appl.
\textbf{298}, 487--500 (2004)

\bibitem{Foka94}  Fokas, A.S.,  Liu, Q.M.:  Nonlinear interaction of traveling waves of nonintegrable equations. Phys. Rev. Lett. \textbf{72},  3293--3296 (1994)

\bibitem {liu-fokas}  Liu, Q.M.,  Fokas, A.S.:
 Exact interaction of solitary waves for certain nonintegrable equations. J. Math. Phys.
 \textbf{37},
 324--345 (1996)


 \bibitem {ch-dav-2011}    Cherniha, R., Davydovych, V.:  Conditional symmetries
 and exact solutions of
 the diffusive Lotka-Volterra system. Math. Comput. Modelling. \textbf{54},
 1238--1251 (2011)


 \bibitem {ch-dav-2012}    Cherniha, R., Davydovych, V.:  Conditional symmetries and exact solutions
 of  nonlinear   reaction-diffusion systems with non-constant
 diffusivities. Commun. Nonlinear. Sci. Numer. Simulat. \textbf{17},
 3177--3188 (2012)

\bibitem{ch-2010}  Cherniha, R.:  Conditional symmetries
 for systems of PDEs:  new definition and its application for
 reaction-diffusion systems. J. Phys. A: Math. and Theor. \textbf{43}, 405207
 (13pp) (2010)


\bibitem {olv}   Olver, P.J.:
 Applications of Lie Groups to Differential Equations. Springer, Berlin (1986)

\bibitem {b-k} Bluman, G.W., Kumei, S.:  Symmetries and Differential Equations. Springer,
Berlin (1989)

 \bibitem{ch-pli-08} Cherniha, R., Pliukhin, O.:  New conditional
symmetries and exact solutions of reaction-diffusion systems with
power diffusivities. J. Phys. A: Math. Theor. {\bf 41}, 185208
(15pp) (2008)

\bibitem{barannyk2002}  Barannyk, T.:  Symmetry and Exact Solutions for Systems
of Nonlinear Reaction-Diffusion Equations. Proceedings of Institute
of Mathematics of NAS of Ukraine. \textbf{43}, 80--85 (2002)

\bibitem{murata-06} Murata, S.:  Non-classical symmetry and Riemann
invariants.  Int. J. Non-Lin. Mech. \textbf{41}, 242-246 (2006)

\bibitem{arrigo2010}  Arrigo, D.J., Ekrut, D.A., Fliss, J.R., Long, Le: Nonclassical symmetries of a
class of Burgers' systems. J. Math. Anal. Appl.  \textbf{371},
 813--820 (2010)

 \bibitem {ch-se-03}   Cherniha, R.,  Serov, M.:
Nonlinear  Systems of the Burgers-type Equations:  Lie and
$Q$-conditional Symmetries,  Ans\"atze and Solutions.
J.Math.Anal.Appl. \textbf{282}, 305--328 (2003)


\bibitem {kingston-91}  Kingston, J.G.:
On  point transformations of evolution equations. J. Phys. A.
 \textbf{24}, 769--774 (1991)

 \bibitem {ch-se-ra-08} Cherniha, R., Serov, M., Rassokha, I.:  Lie Symmetries   and Form--preserving
  Transformations of Reaction--Diffusion--Convection Equations. J. Math. Anal. Appl.
  \textbf{342},
  1363--1379 (2008)

 \bibitem {van-pop-sop-09}  Popovych, R.,  Sophocleous, C.,  Vaneeva,
 O.:
Enhanced group analysis and exact solutions of variable coefficient
semilinear diffusion equations with a power source. Acta Appl. Math.
\textbf{106}, 1--46 ( 2009)

 \bibitem {ch-myr-2010} Cherniha, R., Myroniuk, L.:  Lie Symmetries and  Exact Solutions of  a Class of Thin Film
 Equations.
  J. Phys. Math. \textbf{2},  1--19 (2010)

 \bibitem {ch-dav-2013} Cherniha, R., Davydovych, V.:  Lie and Conditional symmetries
  of the 3-D diffusive  Lotka-Volterra system. J.
Phys. A: Math. Theor. -- 2013. -- Vol. 46. -- 185204 (14 pp).



\end{thebibliography}
\end{document}